\documentclass[pre,superscriptaddress,a4paper,twocolumn,showpacs,amsmath,amssymb]{revtex4}
\usepackage{graphicx}
\usepackage[latin1]{inputenc}

\begin{document}

\title{Dynamics of Semiflexible Polymers in a Flow Field}

\date{26.\ April 2006}
\author {Tobias Munk}
\email{tobias.munk@lmu.de}
\affiliation {Arnold Sommerfeld Center and Center for NanoScience,
  Ludwig-Maximilians-Universität München, Theresienstraße 37, 80333 München,
  Germany} 
\author {Oskar Hallatschek}
\affiliation {Lyman Laboratory of Physics, Harvard University, Cambridge, MA
  02138, USA} 
\author {Chris H. Wiggins}
\affiliation {Department of Applied Physics and Applied Mathematics and Center
  for Computational Biology and Bioinformatics, Columbia University, 500 West
  120th Street, New York, NY 10027, USA} 
\author {Erwin Frey}
\affiliation {Arnold Sommerfeld Center and Center for NanoScience,
  Ludwig-Maximilians-Universität München, Theresienstraße 37, 80333 München,
  Germany}




\begin{abstract}
  We present a novel method to investigate the dynamics of a single semiflexible
  polymer, subject to anisotropic friction in a viscous fluid. In contrast to
  previous approaches, we do not rely on a discrete bead-rod model, but introduce a
  suitable normal mode decomposition of a continuous space curve. By 
  means of a perturbation expansion for stiff filaments we derive a closed set
  of coupled Langevin equations in mode space for the nonlinear dynamics in two
  dimensions, taking into account exactly the local constraint of
  inextensibility. The stochastic differential equations obtained this way are
  solved numerically, with parameters adjusted to describe the motion of actin
  filaments. As an example, we show results for the tumbling motion in shear flow.
\end{abstract}

\maketitle

\section{Introduction}

In the course of evolution, nature found a both amazingly simple and robust way
to sustain the mechanical stability of biological cells, while at the same time
providing them with extraordinary dynamic capabilities, like 
growing, moving, and dividing. The basic structure elements making this possible are
semiflexible polymers, in the cytoskeleton present in form of F-actin,
intermediate filaments and microtubuli. Two characteristic properties
distinguish them from most of the other natural and synthetic polymers: They
possess a certain stiffness which energetically suppresses bending, and they are
to a high degree inextensible, i.e., their backbone cannot be stretched or
compressed. Moreover, electric charge and polarity effects as well as the
ability to assemble and disassemble are essential for the dynamics of the whole
cytoskeleton network. 

In the last 15 years enormous progress has been made in experimental observation
and theoretical description of these physical aspects. To give some examples,
fluorescence video microscopy of labeled filaments has allowed for the detailed
study of statistic properties of F-actin \cite{Kaes:96}, DNA
\cite{MaierRaedler1999}, and microtubules \cite{pampaloni:06}. Quantities like
the end-to-end distance \cite{WilhelmFrey1996} and force-extension relations
\cite{Kroy:95} have been calculated and verified experimentally
\cite{BustamanteBryantSmith2003,LeGoff:02}, the latter being accessible by means
of optical and magnetical tweezers. Furthermore, fluorescence correlation
spectroscopy \cite{LummaEtAl2003,Winkler:06,Shusterman:04} and light scattering
\cite{Kroy:97} have been used to obtain, e.g., the mean square displacement and
dynamic structure factor.

In this paper we concentrate on the theoretical description of the dynamics of a
single semiflexible polymer.  This is the relevant model not only in the dilute
limit, but also essential for the understanding of the medium and high frequency
response of networks of semiflexible filaments \cite{Gardel:04}. Concerning the
theory, the textbook models of Rouse and Zimm \cite{doi:86} have to be extended,
since they apply to Gaussian chains only and thus fail to incorporate the
effects of semiflexibility. A systematic analysis of the dynamics was limited in
this field to the linear regime until recently, when the dynamic propagation and
relaxation of tension could be elucidated
\cite{Bohbot-RavivEtAl2004,Hallatschek:04}, and a comprehensive, unified theory
worked out \cite{Hallatschek:05}. The numerical approach usually adopted for
polymeric systems is to construct a polymer from a finite number of beads, each
of which is connected with two neighbors by a stiff rod or spring
\cite{Bird:87}.  Our goal is to set up a different method more suited to the
subtleties of semiflexibility, such that the numerical description of these
filaments becomes tractable when they are subjected to fluid flow. The purpose
of this work is twofold: First, we want to establish a new method that covers
the above-mentioned goals, and second, the results we present in applying this
technique are directly relevant to experiments with polymers in a shear flow.

Many authors have addressed systems with semiflexible polymers by means of
different bead--rod/spring based techniques
\cite{Ermak:78,Fixman:78,GrassiaHinch1996,Morse:04,Hofmann:00}. However, maintaining the
mechanical constraint of a constant bond length becomes complicated with
increasing resolution, since it couples the motion of all beads. The
controversial question arises whether this requirement has to be implemented as
a literally rigid constraint or by an infinitely stiff potential, since these
two cases differ in their statistical mechanics \cite{vanKampen:84}.  One
often chooses to address an infinitely stiff bead-spring chain by means of a
rigidly constrained system, which is more feasible concerning computational
time. Then, an additional pseudo-potential has to be applied to guarantee the
correct equilibrium Boltzmann distribution \cite{Hinch:93,PasqualiMorse2002}.
Moreover, timescales present major limitations when solving stiff systems
numerically \cite{Everaers:99}. The characteristic relaxation time of a bending
mode imposed on the filament is inversely proportional to the fourth power of
the corresponding wavenumber. The largest wavenumber that can be resolved by
discretized models is proportional to the number of beads. Hence the time step
one has to choose is inversely proportional to the quartic number of monomers,
for the shortest wavelength undulations to be sampled correctly. At the same
time, the longest mode needs to have enough time to relax to equilibrium, and
since these two times differ by the fourth power of the resolution of the
spatial discretization, the necessary computational time can become
prohibitively long.

As a consequence, the numerical approach has been limited, when semiflexible
systems with constraints were to be investigated. Some dynamic properties could
be grasped by a suitable combination of multiple short runs \cite{Everaers:99},
but this is not possible when further external timescales enter, which do not
match the intrinsic limitations. For instance, this is the case when the polymer
is subject to a fluid flow. Corresponding simulations have been reported with a
quite small resolution of nine beads \cite{Montesi:05}. On the other hand,
experiments of this type have already been carried out for DNA
\cite{Perkins:97,Smith:99,Teixeira:05,Schroeder:05,Gerashchenko:06}, and
simulations have been presented to examine their findings
\cite{Hur:00,Schroeder:05b,Puliafito:05}. Indeed, the latter have always been
done with models that allow for a finite or even infinite extensibility of the
chains. This might be a suitable approach for coil--like DNA molecules, but it
would miss important physics, if applied to stiffer filaments like F-actin or
(pre--)stretched DNA \cite{Obermayer:06}.

Recently, a new idea has been presented \cite{wiggins:03} for an alternative
approach that avoids these difficulties for two-dimensional systems. This
concept starts from a continuous model for the semiflexible polymer and uses a
suitable normal mode analysis. Some of these ideas have been utilized before to
describe the deterministic \cite{Becker:01}, linear behaviour of semiflexible
filaments in a viscous solution \cite{wiggins:98}. However, the extension to the
nonlinear case at finite temperature has proven to be subtle, and it is the goal
of this work to consistently establish the approach and explore some of the
possibilities it can offer.

\section{Method}
\label{sec:method}

\begin{figure}[hbt]
  \centering
  \includegraphics[width=.82\columnwidth]{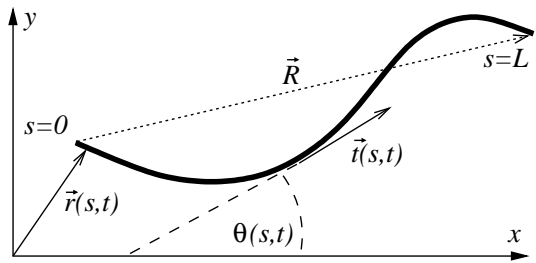}
  \caption{Sketch of a polymer of length $L$, represented by a continuous space curve $\vec r
    (s,t) $. An example is shown for the local unit tangent vector $\hat t(s,t)$,
    as well as the end-to-end distance vector $\vec R$.}
  \label{fig:sketch}
\end{figure}

In our model, a semiflexible polymer is represented by a space curve
$\vec r(s,t)$, parameterised by the arclength $s$
(Fig.~\ref{fig:sketch}).  The minimal model for the description of a
semiflexible polymer in this representation is the wormlike chain
\cite{KratkyPorod1949,SaitoTakahashiYunoki1967}, valid when the
detailed properties on the atomic and monomeric scale are not
important anymore \cite{doi:86}. Then, the hamiltonian for the elastic
energy is given by the integral of the squared curvature $c(s,t)$,
multiplied with the bending modulus $\kappa$,
\begin{equation}
  \label{eq:1}
  \mathcal H_\text{el} =\frac{\kappa}{2}\int_0^L \!ds\,c^2.
\end{equation}
The bending modulus can be expressed in terms of the persistence length
$\ell_p$, the characteristic length for the exponential decay of the
 tangent autocorrelation \cite[§127]{LandauLifschitz:87-statI}: 
\begin{equation}
  \label{eq:bending}
  \kappa=k_BT\ell_p(\mathrm{dim}-1)/2,
\end{equation}
where ``dim'' denotes the dimension of the embedding space.  For synthetic
polymers, the persistence length is typically of the order of the polymer's
diameter, $\ell_p\approx a$. The tangential correlations thus decay very fast,
hence these polymers are called flexible. Semiflexible are polymers with
$\ell_p\gg a$, which is the case for the biopolymers inside the cell. Depending
on the ratio of contour length and persistence length we can distinguish
different regimes: In case of DNA one usually deals with the flexible limit,
$L\gg\ell_p$; in this work however we are exploring the stiff regime, where
$\ell_p\geq L$. The latter matches in nature, e.g., to the properties of F-actin
and microtubuli.

Due to low Reynolds numbers,  the motion of $µ$m-sized objects like
biopolymers in solution is overdamped, i.e., friction exceeds inertia by several
orders of magnitude \cite{FreyKroy:04}. As a consequence, inertia terms can
safely be neglected, and the equations of motion are of first order in time. The
stochastic dynamics of a polymer in a quiescent solvent can thus be described by
the Langevin equation \cite{doi:86,kroy:00}
\begin{equation}
  \frac{\partial }{\partial t}\vec r(s,t) = \int_0^L\!ds'\,\mathbf H[\Delta\vec
  r\,] \left\{ -\frac{\delta\mathcal H[\vec r(s'', t)]}{\delta\vec
    r(s',t)} + \vec\xi(s',t) \right\}.\label{eq:langevin} 
\end{equation}
 Here $\mathbf H[\Delta\vec r\,]$ is the mobility
 tensor, with $\Delta\vec r\,\equiv\vec r(s,t)-\vec r(s',t)$, and $\vec\xi$ the
 noise, which we assume to be Gaussian distributed with mean zero.  The energetic
 part of the Hamiltonian $\mathcal H[\vec r\,]$ is given by $\mathcal
 H_\text{el}$, and we will discuss necessary additions below.

 Concerning hydrodynamics, we adopt the free draining approximation,
 i.e., neglect all nonlocal interactions. This approximation can be
 justified by evaluating the Fourier transformation of the Green's
 function of a hydrodynamic force field (Oseen tensor), which gives
 only a very weak, i.e.\ logarithmic, mode dependence of the mobility
 \cite{Frey:91}. The underlying physical rationale is the mostly
 straight conformation of a stiff filament. For the same reason,
 excluded volume effects can safely be neglected. We implement the local
 friction of the polymer by using the anisotropic mobility of a stiff
 rod, which differs by a factor of two for the motion parallel and
 perpendicular to its tangent vector $\hat t=\partial_s\vec r$. Hence
 $\mathbf H[\Delta\vec r\,]\to\boldsymbol{\mathcal
   P}(r(s,t))\delta(s-s')$, with the local mobility
 \cite{doi:86,Batchelor:70}
\begin{equation}
  \boldsymbol{\mathcal P} = \mu_\perp\hat n\otimes\hat n + \mu_\parallel\,\hat
  t\otimes\hat t\,.\label{eq:Porig}
\end{equation}
Here, $\hat n$ denotes the unit normal vector. The tangent and normal vector are
related by the Frenet-Serret-equations, which in two dimensions read
$\partial_s{\hat  t}=c\hat n$, $\partial_s{\hat n}=-c\hat t$. 
The friction coefficients (per length) are obtained from the solvent viscosity
$\eta$ and polymer diameter $ a$ by 
$\mu_\perp=\frac{1}{2}\mu_\parallel=\ln(L/a)/2\pi\eta$.

Due to the presumed Gaussian nature of the noise we only need to specify the second
moment of $\vec \xi$. By relating Eq.~(\ref{eq:langevin}) to the 
appropriate Smoluchowski equation \cite{gardiner:04} and requiring a Boltzmann
distribution in equilibrium we obtain
\begin{equation}
  \label{eq:2}
  \langle \vec\xi(s,t)\otimes\vec\xi(s',t')\rangle = 2k_BT\boldsymbol{\mathcal
    P}^{-1}\delta(s-s')\delta(t-t'). 
\end{equation}
Based on physical considerations given in appendix~\ref{sec:stoch-integr} we
will interpret the noise according to Ito \cite{vankampen:01}. In fact, some of
the algebra necessary in the following relies on this interpretation.

A further important ingredient to the wormlike chain model is the 
inextensibility of the filament:
We assume the local length to be constant by imposing the constraint 
$ \partial_s\vec r\,^2\equiv\hat t\,^2=1$. To satisfy it, we introduce a
 Lagrange multiplier function $\Lambda(s,t)$ such that the full Hamiltonian
 reads \cite{goldstein:95}
\begin{equation}\label{eq:2b}
  \mathcal H=\mathcal H_\text{el}- \frac{1}{2}\int_0^L\!ds\,\Lambda(s,t) [\partial_s\vec r(s,t)]^2.
\end{equation}
Although the constraint is identically satisfied in the arc\-length
parameterization, we do need a Lagrange multiplier function here to make the
variation of the coordinates in Eq.~(\ref{eq:langevin}) independent of each
other \footnote{For the thermodynamics, a Smoluchowski equation chosen appropriately
leads to a constraint fulfilled exactly \cite{doi:86}. This has to be contrasted
with the thermodynamic approach of Lagrange transformations, which only enforce mean
values.}. In the following, we will be able to solve the equations
of motion perturbatively for the Lagrange multiplier $\Lambda(s,t)$, hence the validity of the
inextensibility constraint is guaranteed locally for all times. Physically,
$\Lambda(s,t)$ corresponds to the tension acting against forces that would
elongate or compress the filament's backbone.

In evaluating the functional derivative of Eq.~(\ref{eq:2b}) we obtain from
Eq.~(\ref{eq:langevin}) the  nonlinear equation of motion  
\begin{equation}
  \label{eq:eom1}
  \dot{\vec r}-\boldsymbol\Gamma\cdot\vec r= \boldsymbol{\mathcal  P}
  \left\{
    -\kappa
      \vec r\,'''' -
      {(\Lambda \vec r\,')}'
    + \vec  \xi
  \right\}.
\end{equation}
On the left hand side we have subtracted the 
incompressible, homogeneous \footnote {A homogeneous flow can be parameterised
  linearily in Cartesian coordinates.} flow $\vec u=\boldsymbol\Gamma\cdot\vec r$ to 
account for the influence of an externally driven flow field. In
Eq.~(\ref{eq:eom1}) and the following, a prime
indicates a derivative with respect to the arclength. 

When constrained systems similar to Eq.~(\ref{eq:eom1}) are approximated by
discrete bead--rod models, one has to introduce a pseudo-potential (also called
metric force) for the system to evolve into the correct equilibrium state
\cite{Fixman:78,Hinch:93,Montesi:05}. In contrast, we use a spectral approach,
in which the arclength dependence is continuous (before explicitly evaluated on
a computer, of course), but we truncate the expansion in terms of a finite
number of wavelengths.  Evaluating the metric determinant by means of the
recursion relation proposed in Ref.~\cite{Fixman:74} leads to contributions
proportional to powers of the spatial resolution. Thus in case of a continuous
representation, we do not need to bother corrections of this kind.

To grasp the essential physics we will from now on consider the
two-dimensional motion of a filament in a three-dimensional embedding
fluid.  This is actually a standard situation for experiments
\cite{MaierRaedler1999,Teixeira:05}, since the interesting dynamics is
often restricted to two dimensions, while still allowing for a
three-dimensional transfer of momentum to the environment.  Note that
this is distinct from systems where the hydrodynamics is confined to
two dimensions.
Thus in the following,
the analysis will be presented for $\text{dim}=2$, cf.\ Eq.\
(\ref{eq:bending}). The tangent and normal vectors then are given by
(Fig.~\ref{fig:sketch}) $\hat t = (\cos\theta,\sin\theta)$, and $\hat
n =(-\sin\theta,\cos\theta)$. To transform the equation of motion to
variables which are scalar, we differentiate Eq.~(\ref{eq:eom1}) with
respect to the arclength and project the result onto the normal and
tangent vector, respectively. Using the Frenet-Serret-Equations, we
arrive at
\begin{align}
  \label{eq:3}
  \begin{split}
    \dot \theta &= \hat n\cdot\boldsymbol\Gamma\cdot \hat t + \mu_\perp \left\{
      -\kappa \left[ c'''-(d+1)\left(c^3\right)'
      \right]\right.\\
    & \hspace{6ex}\left.-(d+1)\Lambda'c-\Lambda c'+ \left[ \hat
        n\partial_s+(d-1)c\hat t \right]\cdot\vec \xi
    \right\},
  \end{split}\\
  \label{eq:4}
  \begin{split}
    0 &= \hat t\cdot\boldsymbol\Gamma\cdot \hat t+\mu_\perp \left\{ -\kappa
      \left[ c^4-3d(cc')'-cc'' \right]\right.\\
      &\hspace{10ex}\left.+c^2\Lambda-d\Lambda''+ \left[ d\,\hat
        t\partial_s+(d-1)c\hat n \right]\cdot\vec \xi \right\}.
  \end{split}
\end{align}
These equations describe the motion of a semiflexible filament in
its center of mass inertia frame, since the information on the absolute
coordinate gets lost in evaluating the additional derivative of
Eq.~(\ref{eq:eom1}). 
The constant $d$ gives the ratio of parallel and perpendicular friction,
$\mu_\perp=\mu_\parallel/d$.  Slender-body hydrodynamics is described by the
anisotropic case ($d=2$), physically corresponding to the difference in drag
between normal and tangential motion of a slender body in Stokes flow.
The simpler isotropic equations are obtained with $d=1$.

\subsection{Normal Mode Analysis}

The leading contribution governing the elastic dynamics of Eq.~(\ref{eq:3}) is
the second term on the right hand side,
$-\kappa\partial_s^3c(s,t)=-\kappa\partial_s^4\theta(s,t)$. For a suitable normal
mode decomposition of the equations (\ref{eq:3}) and (\ref{eq:4}) we thus may
use the eigenfunctions of the biharmonic operator $\partial_s^4$. In this paper,
we consider free boundary conditions $\vec r\,''(0)=\vec
r\,''(L)=0$, $\vec r\,'''(0)=\vec r\,'''(L)=0 $ 
\cite{LandauLifschitz:91-el,WigginsEtAl1998}, corresponding to the situation of
a filament fluctuating freely in flow. In angular coordinates this
translate into 
\begin{equation}
  \theta'(0) =\theta'(L) =0 \text{, and } \theta''(0)=\theta''(L)= 0.\label{eq:5}
\end{equation}
Furthermore, the tension has to vanish at the
boundaries,  $\Lambda(0)=\Lambda(L)=0$. 

The biharmonic operator is not Hermitian with respect to a single set of
eigenfunctions obeying (\ref{eq:5}); however, it is Hermitian with respect to a
\emph{bi}orthogonal set of functions, where the first set $w^\alpha(s) $ obeys
the boundary conditions of Eqs.~(\ref{eq:5}), and the second set $w_\alpha(s) $
satisfies
\begin{equation}
  w_\alpha(0) =w_\alpha(L) =0 \text{, and } w_\alpha'''(0)=w_\alpha'''(L)= 0.\label{eq:6}
\end{equation} 
These functions are solutions  
of the eigenvalue problems $\partial_s^4w^\alpha =
{k_\alpha^4}/{L^4}\,w^\alpha$, $\partial_s^4w_\alpha =
{k_\alpha^4}/{L^4}\,w_\alpha$, respectively, with identical eigenvalues
$k_\alpha^4$, to be found from the solvability condition  $\cos k_\alpha\cosh
k_\alpha=1$. The general solutions $w^\alpha(s) $, $w_\alpha(s) $ are linear
combinations of trigonometric and hyperbolic functions
\cite{LandauLifschitz:91-el}, and a polynomial of third order for the zeroth
eigenvalue $k_0=0 $. They are explicitly given in
App.~\ref{sec:eigenfunctions}.  Some nice experimental snapshots of the 
mode dynamics of F-actin can be found in Ref.~\cite{Vonna:05}.

The completeness of these two sets of eigenfunctions has not been shown
yet. Neither a variational approach \cite{Courant:67} nor a comparison with a
complete basis \cite{Brauer:65} seem to work in our case. However, we
do not regard this as a major issue, since the failure is only due to our
specific nonstandard set of boundary conditions---for other boundary
conditions completeness can be shown \cite{Courant:67}. Furthermore, in order to
implement the spectral approach, we will have to truncate all mode expansions,
anyway. 

We make use of the two sets of eigenfunctions to obtain normal mode
expansions of the angle and tension (latin indices always start from 1, greek
ones from 0):  
\begin{align}
  \theta(s,t) &= \theta_0(t) +
  \epsilon\sum_{j=1}^\infty\theta_j(t)w^j(s),\label{eq:8}\\
  \Lambda(s,t) &=
  \sum_{\nu=0}^\infty\Lambda^\nu(t)w_\nu(s). \label{eq:9} 
\end{align}
In the first line we have separately written the zeroth mode $\theta_0(t)$,
since it is independent of the arclength; it describes the motion of a stiff
rod.  The higher modes include undulations with successively smaller wavelength.
The perturbation parameter $\epsilon$ will be defined by equipartition, as
illustrated below. Equivalently, the zeroth tension mode $\Lambda^0(t)$ gives
the tension distribution in a straight rod.  By using the spectral expansion of
the tension in the second line we assigned two additional boundary conditions to
the tension [rightmost part of Eq.~(\ref{eq:6})]. We do not expect this
vanishing cubic contribution to be a noticeable restriction to the tension at
the edges. However, there is no simple physical interpretation to it.

As a first application of the mode expansions we
insert Eq.~(\ref{eq:8}) into the wormlike chain Hamiltonian (\ref{eq:1}),
which in two dimensions can be written as 
\[\mathcal H_\text{el}  = \frac{\kappa}{2}\int_0^L\!ds\,(\partial_s\theta)^2.\]
By means of the equipartition theorem and
Eq.~(\ref{eq:bending}) we get an expression for the mean size of the mode
amplitudes (cf.\ App.\ \ref{sec:eigenfunctions}), dependent only on the relative
persistence length: 
\begin{equation}
  \label{eq:10}
  \epsilon\sqrt{\langle\theta_j^2\rangle} =
  \sqrt{2\frac{L}{\ell_p}}\frac{1}{k_j}\,.
\end{equation} 
This suggests to define a flexibility parameter $\epsilon=\sqrt{L/\ell_p}$, which is small
for stiff filaments. In the following section we will use $\epsilon$ to set up a
perturbation expansion for the equations of motion. By definition Eq.~(\ref{eq:8}),
all angular modes $\theta_\beta$ are of order 1 with respect to this expansion.

\subsection{Perturbation Expansion}
\label{sec:pert-expans}

The temperature $T$ enters the equation of motion in two ways: First, in the
amplitude of the noise, Eq.~(\ref{eq:2}), and second, in the expression for the
bending stiffness, Eq.~(\ref{eq:bending}). A perturbation theory with respect to
a single independent parameter $\epsilon $ will thus in general lead to
different results, depending on the choice of the dependent parameter, $\kappa$
or $T$. However, a difference only appears in the expressions of cubic order in
$\epsilon$. Since we aim to a solution in second order, this is of no concern
for us. For concreteness, we choose the example of a constant curvature. Then,
$\epsilon$ enters the equations only via the temperature in the noise
correlator. We make all variables dimensionless by $t/B\to t$,
$B=L^4/\kappa\mu_\perp$, $\vec\xi L^3/\epsilon \kappa\to\vec \xi$, $\Lambda
L^2/\kappa\to\Lambda$, $\mu_\perp\boldsymbol{\mathcal
  P}^{-1}\to\boldsymbol{\mathcal P}^{-1}$, insert the expansions (\ref{eq:8}),
(\ref{eq:9}) into the equations of motion and project them on mode $w_\beta$.
The projection, if not evaluated, is abbreviated by
\[ [\ldots]_\beta=\int_0^L\!d s\,w_\beta\ldots.\]
 We furthermore introduce for the flow dependent terms   
\[ \dot \gamma g_\beta^\parallel=[\hat t\cdot\boldsymbol{\Gamma}\cdot\hat
t\,]_\beta/L, \qquad \dot\gamma g_\beta^\perp= [\hat
n\cdot\boldsymbol{\Gamma}\cdot\hat t\,]_\beta/ L,\] 
where $\dot\gamma$ is the
strength of the flow in units of inverse seconds. $g_\beta$ are dimensionless,
$\theta$-dependent functions that have to be expanded to the appropriate order.
In these terms we obtain the following equations of motion:
\begin{subequations}
  \label{eq:14}
  \begin{align}
    \label{eq:15}
    \begin{split}
      \epsilon\,\partial_t\theta_j &= -\epsilon k_j^4\theta_j
      -\epsilon\!\!\sum_{\nu=0,i=1}\!\!\!\Lambda^\nu\theta_i
      \Xi^i_{j\nu}+\epsilon\eta_j^\perp \\
      &\phantom{=}\,\,+B\dot \gamma g_j^\perp+\mathcal O(\epsilon^3),
    \end{split}\\
\partial_t\theta_0 &= 
    -\epsilon\!\!\sum_{\nu=0,i=1}\!\!\!\Lambda^\nu\theta_i
    \Xi^i_{0\nu}+\epsilon\eta_0^\perp+B\dot\gamma
    g_0^\perp +\mathcal O(\epsilon^3),\label{eq:17}\\
    \label{eq:16}
    0 &=
    d\tilde k_\beta^2\Lambda^\beta+\epsilon\eta_\beta^\parallel+B\dot\gamma
    g_\beta^\parallel +\mathcal O(\epsilon^2).
  \end{align}
\end{subequations}
Here, a threefold overlap integral of eigenfunctions is abbreviated by
\begin{multline}
  \label{eq:13}
    \Xi^\alpha_{\beta\nu} = \frac{1}{2}
  \left[
    (d+1)\int_0^L\!d s\,w^\alpha
    \left(
      \tilde k_\nu^2w^\nu w_\beta-\tilde
  k_\beta^2w^\beta w_\nu
    \right)\right.\\
    \left.+ (d-1) k_\alpha^2\int_0^L\!d s\,w_\alpha w_\beta w_\nu
  \right],
\end{multline}
and the projected noise is expressed as
\begin{align}
  \label{eq:12}
  \begin{split}
    \eta_\beta^\perp &= \left[\left( \hat
        n\partial_s+(d-1)(\partial_s\theta)\hat t\, \right)\cdot\vec
      \xi\,\right]_\beta,\\
    \eta_\beta^\parallel &= \left[\left( d\,\hat
        t\partial_s+(d-1)(\partial_s\theta)\hat n \right)\cdot\vec
      \xi\,\right]_\beta.
\end{split}
\end{align}
Note its dependence on $\epsilon$, via $\theta$ and via the unit vectors $\hat
t$, $\hat n $.  Furthermore we would like to mention that $g_j^\perp$ has
no contribution to zeroth order in $\epsilon$, thus all terms of
Eq.~(\ref{eq:15}) are at least linear in $\epsilon$. A linear stability analysis
of the bending modes described by Eq.~(\ref{eq:15}) is obtained by expanding
$g_j^\perp$ \cite{wiggins:03}.  Finally, the dimensionless noise has the second
moment
\[\langle \vec\xi(s,t)\otimes\vec\xi(s',t')\rangle = 
4\boldsymbol{\mathcal P}^{-1}\delta(s-s')\delta(t-t'). \]

Since the tension contribution to Eqs.~(\ref{eq:15}) and (\ref{eq:17}) is always
of first order in $\epsilon$, it is sufficient to expand it to one order less
than the angular equations. For this reason Eq.~(\ref{eq:16}) is an algebraic
equation. To obtain the given expression we have furthermore taken advantage of
the eigenfunction's property $ w_\alpha'' = -{\tilde k_\alpha^2}w^\alpha$ (see
App.~\ref{sec:eigenfunctions}). The tension modes $\Lambda^\beta$ can thus be inserted
directly into the angular equations. 
\footnote{Using Eq.~(\ref{eq:16}) for the case of a vanishing background flow,
  one can calculate the correlation of the tension:
\[\langle\Lambda(s,t)\Lambda(s',t')\rangle\!=
\!\frac{(k_BT)^3\ell_p^2\mu_\perp}{2dL^7}\delta(t-t')\!\sum_{\alpha=0}\!\frac{w_\alpha(s)w_\alpha(s')}{\tilde k_\alpha^2}. \] 
This might give a hint for estimating a reasonable size of the spring constant
of bead-spring systems, when calculating their Brownian dynamics.}

Concerning the noise, one obvious method to simplify Eqs.~(\ref{eq:12}) would be
to evaluate the projection integrals and find an expression for the correlation
of the integrated noise. However, there are some complications in our case. It
turns out that the correlation of the integrated noise is neither diagonal with
respect to the mode projection (subscript $\beta$) nor with respect to the
direction of the vector projection (superscripts $\perp$, $\parallel$).
Furthermore, the order of the nondiagonal corrections is such that they have to
be accounted for in an expansion up to order $\epsilon^2$. Thus a computational
solution would have to include a numerical diagonalization of the conformation
dependent noise in each step. To avoid this we diagonalize the inverse mobility
matrix $\boldsymbol{\mathcal P}^{-1}$ analytically and calculate the arclength
integrals in each time step by numerical means. As a drawback this includes the
necessity of a fast random number generator, but still the implementation seems
to us to be easier and faster.

The eigensystem of $\boldsymbol{ \mathcal P}^{-1}=\hat n\otimes\hat n+\hat
t\otimes\hat t/d$ can immediately be read off, and the corresponding
transformation matrix $\mathbf S$ is used to rotate the noise locally,
$\vec{\tilde\xi}:=\mathbf S\cdot\vec \xi$. For the calculation of the second
moment of the rotated noise we necessarily need it to be of Ito type,
since the matrix $\mathbf S$ is nonanticipating only in this case. We obtain 
\begin{equation}
  \left\langle\vec{\tilde\xi}(t,s)\otimes\vec{\tilde\xi}(t,s)\right\rangle =
  4\boldsymbol{\mathcal P}_D^{-1}\delta(t-t')\delta(s-s'),\label{eq:18} 
\end{equation}
where the diagonalized mobility matrix 
\[\boldsymbol{\mathcal P}_D^{-1}=\mathbf S\cdot
\boldsymbol{\mathcal P}^{-1}\cdot\mathbf S=
\begin{pmatrix} 1 & 0 \\ 0 & 1/d
\end{pmatrix}.\]
 The new stochastic variable $\vec{\tilde\xi}$ simplifies
the stochastic integrals (\ref{eq:12}) of the equations of motion: 
\begin{align}
  \label{eq:19}
  \eta_\beta^\perp &=\int_0^1\!ds \left( d\,{w}_\beta(\partial_s\theta)
    \tilde\xi_2-{w}_\beta'\tilde\xi_1 \right), \\
  \label{eq:20}
  \eta_\beta^\parallel &=-d\int_0^1\!ds\, {w}_\beta'\tilde\xi_2 +\mathcal O(\epsilon).
\end{align}
Using this in Eqs.~(\ref{eq:14}) allows
 us to conveniently solve the coupled stochastic
differential equations  by numerical integration \cite{kloeden:95}. The final
equations \emph{without flow} are a coupled set of \emph{linear} equations
(although of quadratic order in $\epsilon$) with multiplicative
noise. Nevertheless, they cannot be solved analytically, due to the complicated
structure of the coefficients.

\section{Results}

\subsection {Verification Without Flow Field}

\begin{figure}[b]
  \includegraphics[width=\columnwidth]{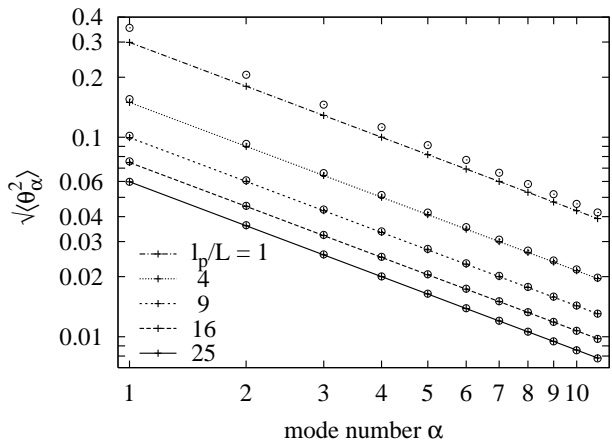}
  \caption{Mean squared amplitudes of normal modes for different persistence
    length. Open circles are numerical results, ``+'' interconnected by lines
    are given by Eq. (\ref{eq:10}). Significant deviations between both are
    visible only for $\epsilon=\sqrt{L/\ell_p}=1$. They are of order
    $\epsilon^3$, terms neglected in the perturbative solution
    Eq.~(\ref{eq:15}) (cf.\ also Fig.\ \ref{fig:2}).} 
  \label{fig:1}
\end{figure}

\begin{figure}[b]
  \includegraphics[width=\columnwidth]{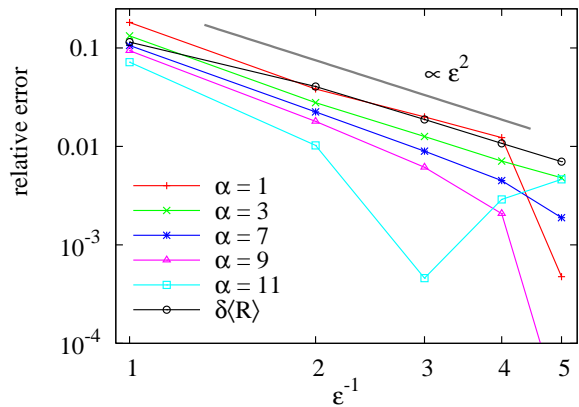}
  \caption{(Color online) Relative errors of some mean square mode amplitudes
    $\epsilon\sqrt{\langle\theta_\alpha^2\rangle}$ (colored) and
     mean end-to-end distance (black circles) versus inverse
    expansion parameter $\epsilon$. Lines are shown to guide the eye. As
    expected from the order of the expansion of 
    the equations of motion, the relative error vanishes proportional to
    $\epsilon^{-2}$, thus the absolute error is $\propto\epsilon^{-3}$.}
  \label{fig:2}
\end{figure}

We have computed equilibrium averages of the squared mode amplitudes to compare
them with the equipartition result, Eq.~(\ref{eq:10}). This is to check the
validity of the approximations we made in the previous section and of the
numerical solution technique. For the first eleven modes shown in
Fig.~\ref{fig:1}, the agreement is excellent in case of $\ell_p/L\geq 9$, i.e.\ 
$\epsilon\leq 1/3$. Slight deviations due to the limited validity of the
$\epsilon$-expansion become visible at $\epsilon=1/3$; in case of $\epsilon=1$,
results differ by about 14\%. To analyze these errors in more detail, we have
plotted the relative differences of analytical and numerical values for the mean
square mode amplitudes versus $\epsilon^{-1}$ in Fig.~\ref{fig:2}. These
relative errors decrease like $\epsilon^2$, as shown by the grey bar. This is
consistent with our second order perturbation expansion, since the first terms
we neglected are of third order, thus their contribution to the relative error
is proportional to $\epsilon^2$.  

In addition to that, Fig.~\ref{fig:2} shows the relative errors of the mean
end-to-end distance $R$ of the polymer. The corresponding exact result is
$\langle R^2\rangle=L^2f_D(L/\ell_p)$, with $f_D(x)=2(e^{-x}-1+x)/x^2$
\cite{KratkyPorod1949,SaitoTakahashiYunoki1967}. The deviations again decrease
as expected proportional to $\epsilon^2$.

\begin{figure}[htb]
  \includegraphics[width=\columnwidth]{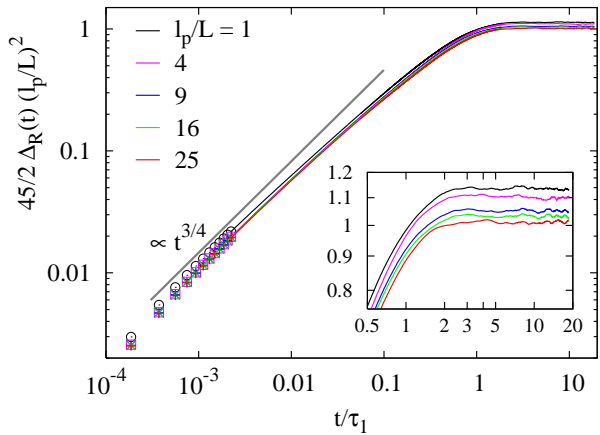}
  \caption{(Color online) Time dependence of the mean square
    displacement $\Delta_R(t)$ of the end-to-end distance. The calculations
    are done in 12-dimensional mode space. Symbols are replaced by
    lines were they would become very dense. The rescaled data for
    different persistence length collapses onto one master curve,
    which behaves $\propto t^{3/4}$ for short times. $\tau_1$ is the
    relaxation time of the longest mode, $\tau_1=k_1^{-4}$, in our units (cf.\
    Eq.~\eqref{eq:15}, and Ref.~\cite{LeGoff:02}). The \emph{inset}
    shows a magnification of the long-time behaviour. The stiffer the
    filament, the closer the numerical results are to the correct
    equilibrium value 1.}
  \label{fig:3}
\end{figure}

To validate dynamic properties, we compare the scaling behaviour of fluctuations
of the mean square end-to-end distance (MSD) $\Delta_R(t)=\langle[R(t)-R(0)]^2\rangle$ with
the analytically known and experimentally verified \cite{Granek1997,LeGoff:02}
behaviour: For short times, $\Delta_R(t)$ grows subdiffusively like $t^{3/4}$, for
long times it approaches the equilibrium value of $\Delta_R(t\to\infty)=2(L/\ell_p)^2/45$.
Our numerical results in Fig.~\ref{fig:3} reproduce this pattern in excellent
quality; effects of the perturbation expansion can only be seen in slight
deviations from the expected plateau value at long times. Comparing
Fig.~\ref{fig:3} to corresponding experimental results (Fig.\ 3 of Ref.\ 
\cite{LeGoff:02}) even shows a similar downturn for times $t/\tau_1<10^{-3}$. In
case of Ref.~\cite{LeGoff:02} this is due to insufficient statistics at small
times because of a limited observation period. In our case, the finite number of
modes accounted for causes a slight suppression of fluctuations at these short
times. \footnote{For very short times, local fluctuations parallel to the
  tangent show a $t^{7/8}$ behaviour \cite{Everaers:99,LeGoff:02b}. However,
  this is a property of terms of the order $\epsilon^3$ and higher in the
  equations of motion, and hence cannot be seen from our calculations.}

To summarize this part, we find that our method 
reproduces all tested equilibrium and dynamic
observables very well within the accuracy expected from the perturbation
expansion. In principle, one could think of even enhancing the accuracy of the
equations. However, in spite of the simplicity of the equipartition expression,
such a correction by means of an additional drift term is quite
involved. This is caused by the strongly coupled nature of the equations, coming
from the Lagrangian constraint and the noise projection. We dismissed such
additions for this work, since the deviations are of minor importance for the
applications we have in mind.

In the following section we will turn to an example of a nonequilibrium system
that can nicely be worked out by means of the technique presented
above. We always present results for filaments with
persistence length $\ell_p/L\geq 4$, thus the systems show the correct
equilibrium behaviour, as demonstrated by now.

\subsection {Motion in Shear Flow}

\begin{figure}[htb]
    \includegraphics[scale=.6]{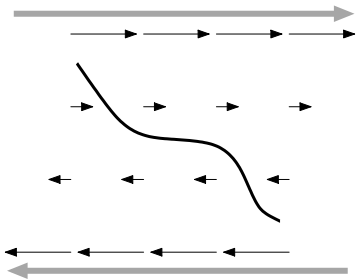}
  \caption{Sketch of a filament in shear flow. The two grey arrows indicate
    walls moving with constant speed, black arrows the velocity of the fluid. At
    the fluid--wall interface, the fluid is transported with the velocity of the
    wall, due to no-slip boundary conditions.}
  \label{fig:shear-sketch}
\end{figure}

A shear flow  is a laminar flow with a velocity field as depicted in 
Fig.~\ref{fig:shear-sketch}. In terms of the velocity gradient matrix
$\boldsymbol\Gamma$ this reads
$\boldsymbol\Gamma=\dot\gamma\big(
\begin{smallmatrix}
  0&1\\0&0
\end{smallmatrix}\big)$, with respect to the coordinate axes of
Fig.~\ref{fig:sketch}.

To gain some understanding of the basics we briefly discuss the case of a stiff
rod exposed to shear. By re-implementing dimensionalized quantities in
Eq.~(\ref{eq:17}) and afterwards taking the limit $\epsilon\to 0 $ we obtain the
equation of motion (cf. Ref.~\cite{Puliafito:05})
\begin{equation}
 \partial_t\theta_0(t)=-\dot \gamma\sin^2\theta_0(t)+\sqrt
 D\eta(t).\label{eq:11}
\end{equation}
Here, the noise is $\delta$-correlated in time, and the diffusion constant
$D=2\mu_rk_BT/L$, with the rotational mobility of a rod $\mu_r=\mu_\perp 12/L^2$
\cite{doi:86}. In the deterministic case, i.e.\ $\eta\equiv 0$, Eq.~(\ref{eq:11}) can be
solved easily and results in a single rotation of the rod, which reaches the
stall line at $\theta=0$ for long times like $t^{-1}$. In terms of stability
analysis the deterministic dynamics corresponds to a flow on a circle
with a half stable fixed point at $\theta=0$. When noise is present, 
$\eta$ is the control parameter of a saddle node bifurcation at $\eta=0$
\cite{Strogatz:00}. The effect of this is that the stochastic forces drive the
rod across the stall line after some time, such that a new rotational cycle
begins. Driven by the shear and the noise, the rod will now rotate again and again, such that
one can e.g.\ measure the rotational times to characterize the stochastic
process. 

For finite $\epsilon$, we have the additional influence of the bending modes.
Apart from shear-induced bending this can also change the behaviour close to the
stagnation line, since filaments with curved conformations might wiggle easier
across this threshold.

\begin{figure}[htb]
  \includegraphics[width=\columnwidth]{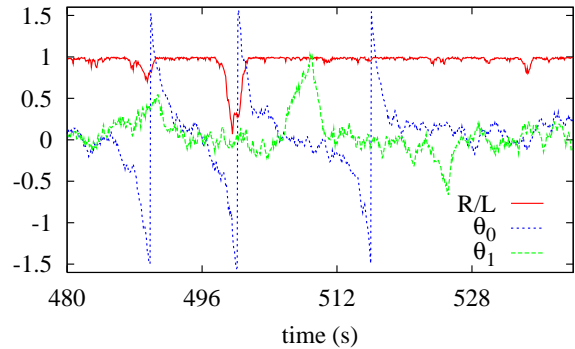}
  \caption{(Color online) Sample trajectories of the
    end-to-end distance $R$ and the first two angular modes, for a flow of strength $\dot\gamma=1.03$/s}
  \label{fig:samples}
\end{figure}

In Fig.~\ref{fig:samples} we show a sample trajectory of a filament with 
relative stiffness $\ell_p/L=6.5$. Here the rotations appear as a flip
of the angular mode $\theta_0$ from $-\pi/2$ to $+\pi/2$. These flips are usually accompanied
by a peak in the first mode $\theta_1$, indicating a sudden  bending event
triggered by the rotation. Furthermore, the end-to-end distance often shows a
short dip corresponding to this intermediate bending.

Extensive experimental \cite{Smith:99,Schroeder:05,Teixeira:05,Gerashchenko:06},
numerical \cite{Hur:00,Montesi:05,Schroeder:05b,Puliafito:05}, and some
theoretical work \cite{Chertkov:05,Puliafito:05} has been presented to
characterize the motion of DNA in shear flow.  However, the relative
persistence length $\epsilon^{-2}$ of DNA is typically about two to three orders
of magnitude smaller than that of F-actin, as already denoted in the beginning
of Sec.~\ref{sec:method}. Consequently, the physics of these two kinds of
semiflexible polymers will be quite different. When rotating in shear, for
example, DNA more or less crawls along itself \cite{Schroeder:05}, whereas
F-actin shows a clear stretch-out in our computations even at intermediate
conformations, i.e., when the end-to-end distance is oriented perpendicular to
the flow direction. Due to the different physics, the numerical methods used for
DNA in the references above are not applicable in case of stiffer filaments like
F-actin. In contrast to those techniques, our method always guarantees the local
inextensibility of the filament.

\begin{figure}[htb]
  \includegraphics[width=\columnwidth]{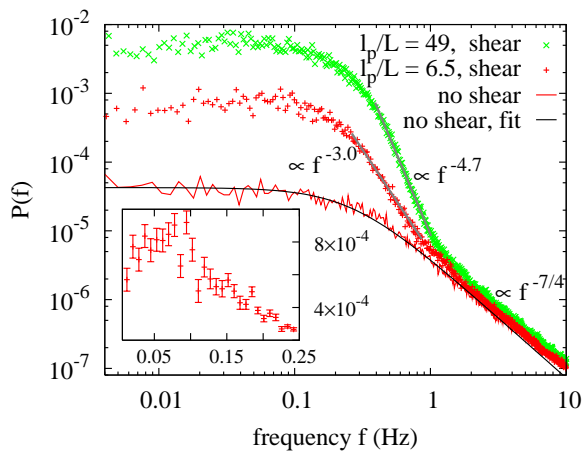}
  \caption{(Color online) Power spectral densities $P(f)$ of the end-to-end distance 
    without flow, and with a shear of strength $\dot\gamma=1.03$ ($\mathrm{Wi}= 0.55$). Green `\textsf x': $\ell_p/L=49$, red `+': $\ell_p/L=6.5$.
    The smooth black curve is a fit of the result in quiescent solvent with the
    appropriate Lorentzian
    \cite{GittesMacKintosh1998},
    used for the plot of the relative PSD
    in Fig.~\ref{fig:5}. With shear, a bump between 0.05 and 0.09 Hz shows the
    characteristic time of rotation of the filament. A different scaling regime
    appears for frequencies smaller than 1 Hz, with an exponent depending on
    properties of the filament. PSD-data for $\ell_p/L=49$ have been multiplied
    by a factor 60 for the purpose of an easier visualization. Calculated with
    10 ($\ell_p/L=6.5)$ and 8 ($\ell_p/L=49$) modes resolution.  \emph{Inset}: Nonlogarithmic version
    of the region around the bump for $\ell_p/L=6.5$, with estimated error bars. The
    number of data points has been reduced by averaging.}
  \label{fig:8}
\end{figure}

A useful observable to get insight into the periodic and stochastic behaviour is the
power spectral density (PSD) $P(f)$, the Fourier transformation of the autocorrelation
function of an observable $x$:
\[ P(f):=\mathcal{FT}\{\langle x(t)x(0)\rangle\}. \]
 In Fig.~\ref{fig:8} we show the PSD of the end-to-end distance $R$ of a
filament with persistence length $\ell_p/L= 6.5 $, both with shear and
without. The strength of the flow is $\dot\gamma=1.03/$s or $\mathrm{Wi}=0.55$,
where $\mathrm{Wi}=\dot\gamma\tau_c$ is the 
dimensionless Weissenberg number, the product of flow rate and 
characteristic relaxation time $\tau_c$ of the system. For $\tau_c$ one may choose the
mean exponential decay rate of the autocorrelation function of the end-to-end
distance. In terms of the dimensionless formulation chosen in
Sec.~\ref{sec:pert-expans}, the bending stiffness $\kappa$ is constant, and only
temperature varies when changing the stiffness parameter $\epsilon$. Thus in this
formulation the relaxation rate is identical for all $\ell_p$. Our data result in a mean of
$\tau_c= 0.53\pm 0.02$\,s, so flow rates $\dot\gamma $ have to be multiplied by
this factor to obtain them in Wi-units.

Without shear, the PSD shows a characteristic Lorentz-like behaviour, where
the only timescale in the system separates the long time plateau from the short
time decay. In case of the end-to-end distance $R$, the short time regime
obeys the power law $P(f)\propto f^{-7/4}$ 
\cite{GittesMacKintosh1998},
which immediately follows from the mean
square displacement $\propto t^{3/4}$. With shear, there is first of all a
pronounced increase in correlations at small frequencies, indicating stronger
long-time correlations due to periodic tumbling events. A shallow bump
appears in the region between 0.05 and 0.09 Hz, indicating a typical frequency
of rotation for the 
given flow strength. This bump is visible more clearly in Fig.~\ref{fig:5},
where we show the ratio of the PSDs with and without flow. 

Coming back to
Fig.~\ref{fig:8}, we identify an additional timescale in the decay for
frequencies below 1\ Hz, separating the high frequency power law
$f^{-7/4}$ from a regime with an exponent with a larger absolute value. This
intermediate sharp decay 
arises due to a subtle interplay between thermal fluctuations and frictional
driving of the shear flow \cite{Hur:00,Schroeder:05b}. However, the power law of
this decay is not generic---it depends on $\ell_p$ in a nontrivial
manner. The detailed study of this phenomenon is deferred to a
later publication.

In the experiments with DNA cited above, the end-to-end distance itself could not
be measured due to the limited optical resolution; instead, the molecular
extension was recorded, as measured by the mean projected extension in flow
direction. However, it is reported that this observable does not show a typical
frequency for a deterministic cycle associated with the tumbling motion in flow,
in contrast to our results for the end-to-end distance of F-actin. We conjecture
that this relatively weak effect might also be connected to the inextensibility
of the filament.

\begin{figure}[htb]
  \includegraphics[width=\columnwidth]{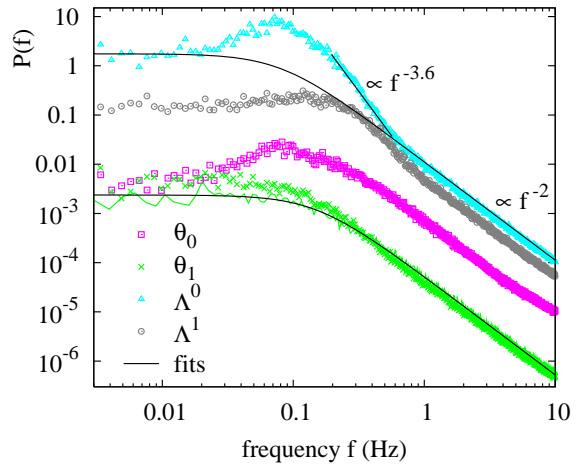}
  \caption{(Color online) Power spectral density of tension modes $\Lambda^0$,
    $\Lambda^1$, and angular  modes $\theta_0$, $\theta_1$. Symbols give
    results with a shear flow of strength 
    $\dot\gamma=1.03/s$ ($\mathrm{Wi} = 0.55$).
    The light line (green online) shows the behaviour of $\theta_1$ without 
    flow, smooth black curves are fitted Lorentzians, as explained in
    the text. 
    A peak in $\theta_0$ and $\Lambda^0$ around 0.5 Hz is a sign
    of the tumbling time of the filament. Concerning $\theta_1$, shear still increases
    correlations at low frequencies.} 
  \label{fig:4}
\end{figure} 

Fig.~\ref{fig:4} shows PSDs of the first two modes for the angle and tension.
In the absence of flow, the stiff filament mode $\theta_0$ is to first order
given by the rotational diffusion of a stiff rod. This leads to a $f^{-2} $
decay in the PSD with nonperiodic boundary conditions (not shown), which still
gives the high frequency regime of the results in shear.  The bumps of
the PSDs of $R$ and $\theta_0 $ are located at about the same frequency---in this
regard it is interesting to note that the end-to-end distance is
\emph{independent} of $\theta_0 $.

\begin{figure}[htb]
  \includegraphics[width=\columnwidth]{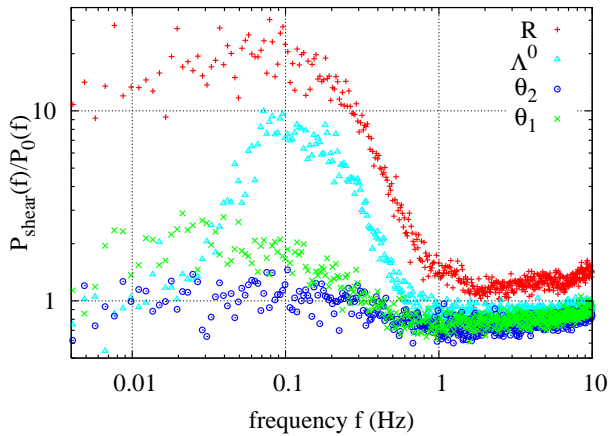}
  \caption{(Color online) Ratio of power spectral density $P(f)$ with shear
    flow to that without flow, for the data of Figs.~\ref{fig:8} and
    \ref{fig:4}. 
    Correlations of $R$ are increased by a factor of more than
    10 at frequencies of 0.3 Hz and below, with a peak between 0.05 and 0.09 Hz. Similarly,
    correlations of the first angular mode $\theta_1$ are pronounced by a
    factor of 2 below 0.1 Hz, and slightly decreased around 1 Hz. This is a
     signature of the semiflexible nature of the polymers under
    investigation, since it refers to the buckling events occurring periodically
    during their rotation in shear. Concerning the
    second angular mode $\theta_2$ (not shown in Fig.~\ref{fig:4}), there is still
    a change from decreased to increased correlations visible when going from
    $f=1$ Hz to 0.1 Hz and below. The tension mode $\Lambda^0$ shows a strong
    tenfold peak at 0.1 Hz when compared to a Lorentzian which has been fitted
    to the generic short- and long-time behaviour.}
  \label{fig:5}
\end{figure}

In a quiescent solvent, Eqs.~(\ref{eq:14}) give Ornstein-Uhlenbeck behaviour for
the higher angular modes $\theta_i$, with small deviations due to the coupling
of the equations and the multiplicative noise. In Fig.~\ref{fig:4} we only show
the first mode $\theta_1 $, where the case without flow has been fitted to a
Lorentzian, whose coefficients differ from the Ornstein-Uhlenbeck process by at
most some percent. The differences between observables recorded with and without
flow are better visible in Fig.~\ref{fig:5}, where we identify an increase below
0.3 Hz and a slight decrease for higher frequencies. The increase of the plateau
for $\theta_1$ is a characteristic of filaments with a relative persistence
length of the order one: Stiffer filaments do not buckle at all during a
rotation, and floppier ones show a different behaviour for strong flows
(crawling near the stall line), or thermally fluctuate so heavily that a
buckling due to shear cannot be identified. The next mode $\theta_2$ still shows
a similar behaviour, but in an already much weaker manner.

The tension modes finally only show noise fluctuations around zero when no flow
is present (not shown), which is obvious from Eq.~(\ref{eq:16}). In shear flow the
power spectrum changes towards a behaviour as characteristic to Lorentzian
curves for low and high frequencies.  This is caused by the fact that there is a
strong linear coupling between each tension mode $\Lambda^\beta$ and its
corresponding angular mode $\theta_\beta$, visible by expanding the flow
dependent part of Eq.~(\ref{eq:16}). The intermediate frequency region of the
PSDs of the tension modes clearly shows signs of driving by shear: A
bump, indicating a typical correlation time, followed by a sharp decay. These
properties are even more pronounced in the tension modes, when comparing to the
angular mode with the same index, respectively. In case of $\Lambda^0$ we
identify a strong peak at the very same position of maximum height already
mentioned for the PSDs of $\theta_0 $ and $R $, and a sharper power law decay
$f^{-3.6}$ towards higher frequencies. We conclude from this that the
periodicity of the tumbling events can be found directly in the autocorrelation
of the tension modes. This is very reasonable, since it follows that a rotation
in shear triggers specific frictional forces acting on the backbone of the
polymer, which have to be withstand by the constraining forces. In
Fig.~\ref{fig:5} we show the PSD of $\Lambda^0 $ relative to a Lorentzian fitted
to the generic short- and long-time behaviour to demonstrate the flow specific
peak.

\subsection {Statistics of Tumbling in Shear}

A further point of interest is the mean time it takes
the polymer to rotate. Experiments with DNA report a power law increase of the
tumbling frequency with shear strength proportional to $\mathrm{Wi}^{0.67}$, confirmed by
appropriate simulations  and scaling analysis \cite{Schroeder:05}. Furthermore,
analytic results for Brownian rods in strong shear consistently give
$\mathrm{Wi}^{2/3}$ \cite{Puliafito:05}. 

\begin{figure}[htb]
  \includegraphics[width=\columnwidth]{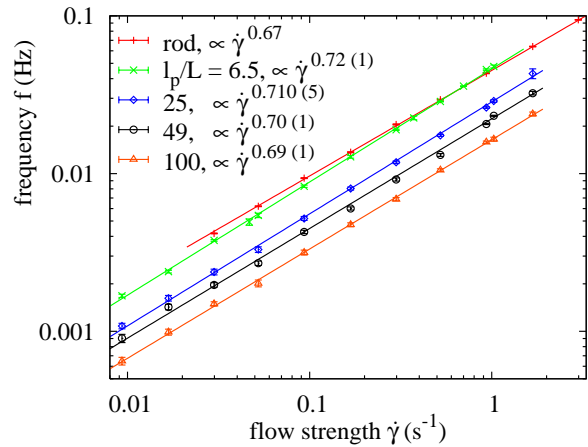}
  \caption{(Color online) Tumbling frequency of semiflexible polymers with different
    persistence length versus strength of shear flow. The change of the power
    law form $\dot\gamma^{0.72}$ at $\ell_p/L=6.5$ towards $\dot\gamma^{2/3}$ for a stiff rod
    indicates a crossover behaviour between rods and flexible polymers.  The
    amplitude shift appears due to the parameterization we have
    chosen (cf.\ Sec.\ \ref{sec:pert-expans}) and has not been removed for
    reasons of readability.} 
  \label{fig:flip}
\end{figure}

Fig.~\ref{fig:flip} shows our results for the mean rotating frequency of
filaments with various persistence lengths. We defined a flip to be a \emph{single}
half turn of an angle $\pi$. As a check of consistency, the frequency of rotation of
a filament corresponding to Figs.~\ref{fig:8} and \ref{fig:4} is 
$0.066\pm 0.002$\,Hz,
which lies very well within the observed peak width of these figures. 

The maximum flow strength for which we can obtain results numerically is limited
by the $\epsilon$-expansion we have exploited in Sec.~\ref{sec:pert-expans}.
Since this expansion requires a small curvature, it breaks down when a powerful
flow strongly bends the filament. For comparison we also plotted numerical
results for a rotating stiff rod, which obeys the expected $\dot\gamma^{2/3}$
power law. The exponent shows a slight crossover behaviour in the semiflexible
regime, in deviation from both the stiff and the flexible limit: As an example,
the rotational frequency of a filament with persistence length $\ell_p/L=6.5$
scales like $\dot\gamma^{0.72\pm 0.01} $. The data of Fig.~\ref{fig:flip}
furthermore demonstrate that the exponent becomes closer to 2/3 the stiffer the
filament, as expected from the stiff rod limit. Note that this relatively weak
crossover effect does not vanish $\propto\epsilon^3$, and thus seems not to be
caused by deviations due to the perturbation expansion.

\section{Conclusion}

We have presented a newly developed method for describing the dynamics
of single semiflexible filaments in a viscous solution, subject to
external flow fields.  Due to the mainly elongated conformations,
hydrodynamic backflow effects are marginal and thus the dynamics can
be formulated in the free draining approximation.  In contrast to
previous approaches based on bead-rod/spring models in real space we
have adopted a spectral method of the equations of motion. A further
simplification can be achieved upon using an angular representation of
the polymer conformations. This has the advantage that there is no
approximation in the bending energy. All of the above allows us to
give an efficient computational approach for calculating Brownian
trajectories for the polymers, and to include nonlinear effects of the
environment without inherent limitations.

The computational time necessary for these numerical solutions is proportional
to the fourth power of the mode number, and linear in the spatial resolution of
the noise along the space curve. The main advantages regarding numerics are,
compared to bead--rod models, first, mode dynamics offers a natural approach to
the long-time dynamics, since the high wavenumber fluctuations are increasingly
irrelevant (cf.\ Fig.\ \ref{fig:1}). Second, the additional time necessary to
satisfy the local constraint of constant length is very small. One reason for
this is that no pseudo--potentials have to be calculated. Finally, one could
even enhance the speed by adapting time and spatial resolution appropriately to
each individual mode, a technique we did not apply, yet. In summary, we are able
to monitor conformational properties very precisely, while taking account to all
modes that give essential contributions to the dynamics of stiff semiflexible
filaments.

Conceptually, our approach should be understood as being complementary
to bead-rod and bead-spring models. In some cases the latter are
advantageous, e.g.\ when dealing with dense solutions of flexible polymers, such that excluded
volume effects are important.

Quantitative tests of the numerical solution show that the method correctly
describes the fluctuations of semiflexible filaments in a quiescent solvent. We
have demonstrated that our technique is capable of describing the long time dynamics in
a sheared environment within reasonable computational time. For the
first time we have calculated power spectral densities and mean rotational
frequencies of F-actin in shear, taking into account exactly the local
inextensibility constraint. The results have enabled us to point on two
crossover phenomena still to be described in more detail. For the future,
extensions of the method should be possible to include e.g. charge effects or 
different boundary conditions. A more challenging task is to extend the analysis
to 3D.

An experimental test of the behaviour of Actin filaments in shear or
elongational flow has not been reported, to our knowledge. However,
the setup used for sheared DNA \cite{Smith:99,Schroeder:05} might also work for
F-actin, and elongational flows are constructed easily by means of microfluidic
techniques \cite{Koester:05}. Thus experiments could in principle be possible
without too much technical effort.

\begin{acknowledgments}
  We are indebted to Thomas Franosch and Abhik Basu for very valuable
  discussions. Moreover, we would like to acknowledge financial support from the
  Intern.\ Graduate School ``Nano-Bio-Technology'' of the Eli\-te\-netz\-werk
  Bayern (TM), the German Academic Exchange Programme DAAD (OH), and the DFG
  through grant SFB 486 (EF).
\end{acknowledgments}

\appendix

\section{The Stochastic Integration}
\label{sec:stoch-integr}

Stochastic equations are of Stratonovich type when the white noise they imply
constitutes an approximation to a noise with finite correlation time. This seems
to apply to us, since physically realistic random forces are never completely
uncorrelated. However, this is not the only criterion for the decision, we also
have to grasp correctly the physical relationship between the 
stochastic variable $\vec r$ and the noise $\vec \xi $ present in our specific
system \cite{vankampen:01}. The term we have to decide about is the mobility
$\boldsymbol{\mathcal P} $, since it multiplies the noise in
Eq.\ (\ref{eq:eom1}). Its physical meaning is to separate locally the velocity of
the filament into a component parallel to its local tangent and another
perpendicular to it. This separation will always refer to the current
conformation and velocity of the space curve \emph{at that very moment of time};
it will be unaffected by any stochastic forces in the future. This amounts to
the definition of a nonanticipating function \cite{gardiner:04}, and is
equivalent to the demand to interpret the noise according to Ito.

\section{Eigenfunctions}
\label{sec:eigenfunctions}

The normalized biharmonic eigenfunctions obeying the boundary conditions of
Eqs.~(\ref{eq:5}) and (\ref{eq:6}) are 
\begin{align}
  w^0 &= 1\,,\label{eq:41}\\
  \label{eq:143}
  \begin{split}
    w^i &= \frac{\cos k_i-\cosh k_i}{\sin k_i-\sinh k_i}
    \left(\cos \frac{k_i}{L} s+\cosh \frac{k_i}{L} s\right)\\
    &\phantom{=}\;+\sin \frac{k_i}{L} s-\sinh \frac{k_i}{L}
    s\:,
  \end{split}\\
w_0 &= 6\frac{s}{L}\left(1-\frac{s}{L}\right)\,,\label{eq:95}\\\label{eq:42}
\begin{split}
  w_i &= \frac{\cos k_i-\cosh k_i}{\sin k_i-\sinh k_i}
  \left(\cos \frac{k_i}{L} s-\cosh \frac{k_i}{L} s\right)\\
  &\phantom{=}\;+\sin \frac{k_i}{L} s+\sinh \frac{k_i}{L} s \,.
\end{split}
\end{align}
They are biorthonormal,
\begin{equation}\label{eq:7}
  \int_0^L \!d s\, w^\alpha w_\beta = L\delta^\alpha_\beta.
\end{equation}

A useful property of these eigenfunctions is
\[{w^\alpha}'' =-{k_\alpha^2}/{L^2}\,w_\alpha, \qquad
 w_\alpha'' = -{\tilde k_\alpha^2}/{L^2}\,w^\alpha.\] 
Here, $\tilde k_0^2 \equiv 12$, and 
$\tilde k_i\equiv k_i$.


\end{document}